\documentclass[prb,preprint]{revtex4}
\usepackage{amsmath}  
\usepackage{amsfonts} 
\usepackage{graphicx} 
\providecommand{\U}[1]{\protect\rule{.1in}{.1in}}
\begin{document}
\title{Stirling engine operating at low temperature difference}
\author{Alejandro Romanelli}
\altaffiliation{alejo@fing.edu.uy}
\affiliation{Instituto de F\'{\i}sica, Facultad de Ingenier\'{\i}a\\
Universidad de la Rep\'ublica\\ C.C. 30, C.P. 11000, Montevideo, Uruguay}
\date{\today}
\begin{abstract}
The paper develops the dynamics and thermodynamics of Stirling engines that run with temperature
differences below $100~^0$C. The working gas pressure is analytically expressed using an
alternative thermodynamic cycle. The shaft dynamics is studied using its rotational equation of
motion. It is found that the initial volumes of the cold and hot working gas play a non-negligible
role in the functioning of the engine.
\end{abstract}
 \maketitle
\section{Introduction}
In the field of energy efficiency, the use of waste energy is one of the keys to improve the
performance of facilities, whether industrial or domestic. In general the waste energy arises as
heat, from some thermal process, that it is necessary to remove. Therefore the use of the waste
energy is usually conditioned by the difficulty of converting heat into other forms of
energy.\cite{tesis4,tesis3} The Stirling engines, being external combustion machines, have the
potential to take advantage of any source of thermal energy to convert it into mechanical energy.
This makes them candidates to be used in heat recovery systems.

The Stirling engine is essentially a two-part hot-air engine which operates in a closed
regenerative thermodynamic cycle, with cyclic compressions and expansions of the working fluid at
different temperature levels.{\cite{Reid,Sier}} The flow of the working fluid is controlled only by
the internal volume changes;  there are no valves and there is a net conversion of heat into work
or vice-versa.

A Stirling-cycle machine can be constructed in a variety of different
configurations.\cite{Walker,Darlington}
In this work we focus on the study of low temperature difference (LTD) Stirling engines, that is,
operating on a temperature difference below $100~^0$C and, in general, in a Gamma configuration.
These engines use a heat source that excludes direct combustion, which occurs at temperatures of
several hundred degrees. The temperature range where the Stirling engine works determines its
geometry and proportions.\cite{Senft3} Stirling engines, with a high temperature difference need a
relatively long distance between the hot and cold chambers to avoid an excessive heat loss between
the chambers, while the size of the heating and cooling surface area is less critical. On the other
hand, LTD Stirling engines require gas chambers with a large surface area to facilitate the heat
transfer with the environment, but there is also less heat conduction from the hot to the cold
chambers so the distance between them can be smaller.

The first reference to LTD Stirling engines is related to the work developed by Ivo Kolin of the
University of Zagreb during the $1970$s and $1980$s.\cite{Kolin1,Kolin2} Subsequently James R.
Senft of the University of Wisconsin also developed engines capable of operating with a
LTD.\cite{Senft3,Senft1,Senft2} These two pioneers show what a sustained work of research and
development can do with a concept: their engines that initially operated at temperature difference
of $44~^0$C evolved to run at only $0.5~^0$C.

At the present time research on Stirling engines is one of the lines that contribute both to the
rational use of energy and to sustainable development.\cite{Kong,Barreto,Jokar} In particular the
solar thermal conversion systems based on these engines are amongst the most interesting and
promising research lines. \cite{tesis4,tesis3,tesis1,tesis2,tesis5,tesis6}

To achieve an adequate theoretical description of the Stirling thermodynamic cycle it is necessary
to adopt certain simplifications. Usually this cycle is modeled by alternating two isothermal and
two isometric processes.\cite{Zeman} In previous papers we developed an alternative approach to the
Stirling cycle that provides analytical expressions for the pressure and temperatures  of the
working gas, and the work and heat exchanged with the reservoirs. The theoretical pressure-volume
diagram achieved a closer agreement with the experimental one than the standard
analysis.\cite{Romanelli1,Romanelli2} Due to the generality of the analytical expressions obtained
they can be adapted to any type of Stirling engine. In the present paper we use the mentioned model
to study the dynamics and thermodynamics of a particular Stirling engine that runs at a low
temperature difference.

The paper is organized as follows. In the next section a simple model of LTD Stirling engine is
presented. In the third section we introduce some aspects of our alternative thermodynamics for the
Stirling engine. In section four we introduce the dynamics of the engine in order to complete the
model and make some numerical calculation. In the last section we present the main conclusions.

\begin{figure}[h]
\begin{center}
\includegraphics[scale=0.5]{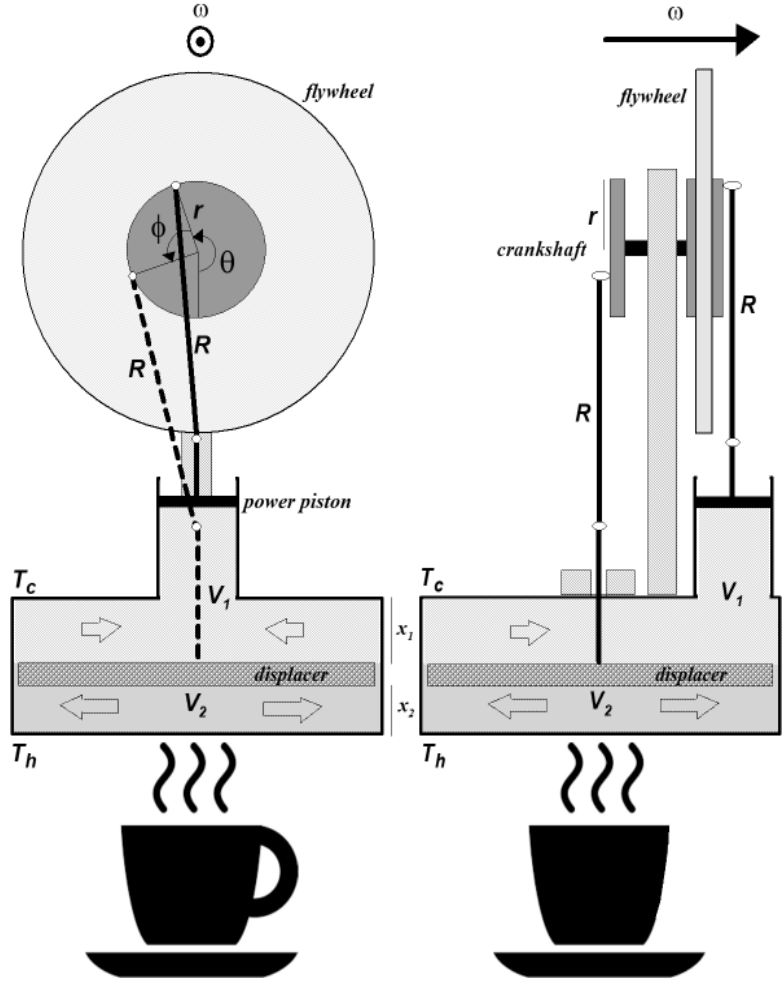}
\end{center}
\caption{Schematic picture of LTD Stirling engine from two perpendicular points of view. $V_1$ and
$V_2$ are the cold and the hot volumes respectively, $x_1+x_2=2r$. Open arrows indicate the
movements of the gas. The flywheel moves counterclockwise and  (in the instant photo) both power
piston and displacer are descending. Several experimental models, whether homemade or commercial,
found in the Internet fit in this scheme.} \label{f1}
\end{figure}
\section{Simple model of a Stirling engine}
In this section we present a simple model of LTD Stirling engine that corresponds to a Gamma
configuration. Figure~\ref{f1} provides a useful insight from head-on and side of the engine
running with the power piston modifying the gas volume and the displacer, continually sweeping the
gas up and down. This configuration has two cylinders joined to form a single connected space with
the same pressure. The cylinder with the larger cross-section contains the displacer and the other
the power piston. The piston and the displacer are joined to a shaft through rods which move in
parallel planes with a relative phase of $\phi\equiv\pi/2$ between the cranks. The shaft has also a
flywheel.

The engine is on the top of a cup of boiling water and the remainder of the engine is in contact
with the environment at room temperature. Note that the engine also would operate on the top of a
cup of cold water, but in this case the flywheel would rotate in the opposite direction. The gas
pressure remains uniform during the entire operation but it changes due to the motion of the piston
and the action of the heat reservoirs, represented by the cup of boiling water and the room
atmosphere.

In the total gas volume two zones, separated by the displacer, can be distinguished by their
temperatures. Our mathematical model assumes uniform temperatures for these zones. This assumption
becomes reasonable when the characteristic times associated with the movement of the pistons are
much larger than those associated with the mean free path of the gas molecules. The hot zone is
below the displacer  where the gas is in contact with the hot reservoir at the external temperature
$T_h$, and the cold zone is above the displacer where the gas is in contact with the cold reservoir
at the external temperature $T_c$. However both zones have the same pressure $P$ because they are
connected through the loose fit of the displacer with its cylinder. When the displacer is in its
lowest position all the gas is in the cold zone, but in the rest of the cycle, the gas is never
completely in either the hot or cold zone of the engine. The piston is also subjected to the
atmospheric pressure $P_0$ from the outside.

In order to start the engine we must set in motion the flywheel externally  and immediately let it
move by the action of the engine. Then the mechanism of energy transfer between the hot and cold
zones begins to operate and it can be qualitatively understood as follows. When the gas is swept by
the displacer to the hot zone it expands, the gas pressure increases and the piston pushes up. When
the gas is swept by the displacer to the cold zone it contracts, and the momentum of the machine,
usually enhanced by the flywheel, pushes the piston down to compress the gas. The phase difference
between displacer and piston is a crucial parameter for the efficiency and power delivered by the
engine. The optimum phase is around $\phi=\pi/2$, as it has been shown.\cite{Senft3,Romanelli1}

The piston and the displacer are connected to the shaft by rods of length $R$ as shown in
Fig.~\ref{f1}. The piston and displacer cylinders have cross-section areas $a$ and $A$
respectively, and the crank has a rotating radius $r$. The relation between the cold and hot gas
volumes, $V_1$ and $V_2$, and the flywheel angle $\theta$ can be obtained from the geometric
analysis of Fig.~\ref{f1}, \emph{i.e.}
\begin{equation}
{V_2}=\left\{1+z\left[1-\cos(\theta+\phi)\right]-\sqrt{1-z^{2}\sin^{2}(\theta+\phi)}\right\}RA,
\label{v2g}
\end{equation}
\begin{equation}
{V_1}=2rA-{V_2}+{V_p}, \label{v1g}
\end{equation}
where $z={r}/{R}$ and $V_p$ is the volume swept by the piston run:
\begin{equation}
{V_p}=\left\{1+z\left(1-\cos\theta\right)-\sqrt{1-z^{2}\sin^{2}\theta}\right\}Ra. \label{v0g}
\end{equation}
The total volume of gas $V$ is given by
\begin{equation}
V={V_1}+{V_2}=2rA+{V_p}. \label{v}
\end{equation}
\section{Thermodynamics of an LTD Stirling engine}
\begin{figure}[ht]
\begin{center}
\includegraphics[scale=0.35]{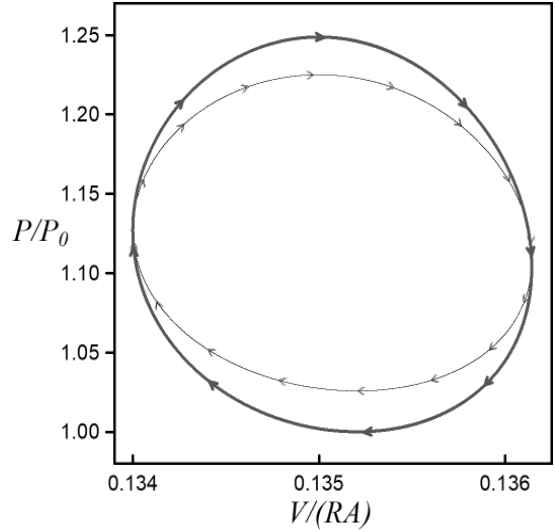}
\end{center}
\caption{Dimensionless pressure-volume diagram for the initial conditions $\theta_0=1.46~\pi$ (red
thick curve). The thin black curve, corresponding to the initial condition $\theta_0=0.46~\pi$, has
been moved in the diagram to the current position for the sake of comparison. The arrows indicate
the evolution with growing $\theta$, and the parameters are: $\phi=0.5~\pi$, $\alpha=1.25$,
$\beta=1.23$, $a/A=0.016$ and $z=0.067$.} \label{f2}
\end{figure}
In previous work we have obtained an analytical generic solution for the pressure and temperatures
inside of any Stirling engine.\cite{Romanelli1} That follows from the application of the first law
of thermodynamics to an ideal gas subjected to polytropic processes.
We have used the mentioned solution to study a particular type of Stirling engine known as
Fluidyne.\cite{Romanelli2} Here we use the expression of the gas pressure as a function of the
volumes (Eq.~(20) of Ref.~\onlinecite{Romanelli1}), \emph{i.e.}
\begin{equation}
P=P_{0}\left(\frac{\alpha\,V_{10}+V_{20}}{\alpha\,V_{1}+V_{2}}\right)\left(\frac{V_{0}}{V}\right)^{\beta-1},
\label{pe}
\end{equation}
where $\alpha$ is the ratio of reservoir temperatures:
\begin{equation}
\alpha\equiv{T_{h}}/{T_{c}}, \label{alfa1}
\end{equation}
$V_{10}$, $V_{20}$ and $V_0$ are the initial values of $V_{1}$, $V_{2}$ and $V$ respectively
(Eqs.~(\ref{v2g}), (\ref{v1g}) and (\ref{v})), and $\beta$ is a parameter called polytropic
index.\cite{romanelli} The typical values of $\beta$ are such that $1\leq\beta\leq \gamma$ with
$\gamma=c_p/c_v$ the quotient of the specific heats of the gas at constant pressure $c_p$ and
constant volume $c_{v}$. For the isothermal and adiabatic cases, $\beta=1$ and $\beta=\gamma$
respectively, Eq.~(\ref{pe}) reduces to the well known solutions found in the specialized
literature of Stirling engines.\cite{Schmidt,Bercho,Formosa}

Equation ~(\ref{pe}) depends on $\alpha$ and it has a finite asymptotic limit when
$\alpha\rightarrow\infty$. This means that after a certain value of $\alpha$, no matter how much we
increase the temperature ratio the pressure and the absorbed heat are bounded, and this in turn
explains why the useful work and the efficiency in any Stirling engine are asymptotically bounded\cite{Romanelli1}.

The pressure-volume diagram of the LTD Stirling engine is obtained starting from Eqs.~(\ref{v0g}),
(\ref{v}) and (\ref{pe}) and implementing numerical calculation. Such diagrams are shown in
Fig.~\ref{f2} for two characteristic initial values of $\theta$ which, as we shall see, correspond
to the maximum and minimum engine work. The area enclosed by the curves represents the total work
$W$ of the cycle, whose expression is
\begin{equation}
W=\oint P\,dV=\int_{\theta_0}^{\theta_0+2\pi}P{\frac{dV}{d\theta}}\,d\theta
=\int_{0}^{2\pi}P{\frac{dV}{d\theta}}\,d\theta, \label{work}
\end{equation}
where $\theta_0$ is the initial angular position that determines the initial conditions for the
cold and hot gas volumes. The last equality in Eq.~(\ref{work}) is obtained with a change of
variable that contemplates the fact that the integrand is a periodic function of $\theta$.  $W$ is
the work available for overcoming mechanical friction losses and for providing useful power to the
engine shaft. $W$ depends on $\theta_0$ because $P$ depends on $\theta_0$ through $V_{10}$,
$V_{20}$ and $V_{0}$ (see Eq.~(\ref{pe})). Here we point out that this dependence, as far as we
know, has not been explored experimentally and it would be interesting to do so.
\begin{figure}[ht]
\begin{center}
\includegraphics[scale=0.35]{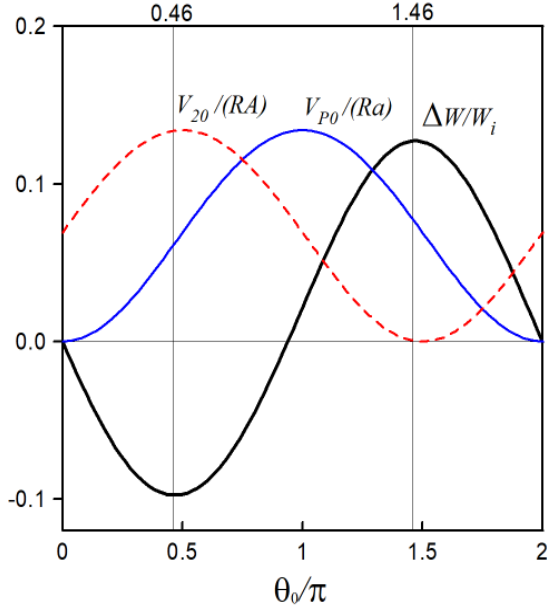}
\end{center}
\caption{Relative work per cycle as a function of the initial angular
condition $\theta_0$, in black thick line, $\Delta W=W-W_i$ where $W_i$ is the work associated with $\theta_0=0$.
The dimensionless initial gas volume of the piston cylinder is shown with the blue thin line. The
dimensionless initial gas volume of the hot zone, is shown with the red dashed line. The values of
the parameters are the same as in Fig.~\ref{f2}.} \label{f3}
\end{figure}
We have integrated numerically Eq.~(\ref{work}) with Eqs.~(\ref{pe}), (\ref{v}) and (\ref{v0g}) as
functions of $\theta_0$, using the standard Simpson's rule. Figure~\ref{f3} shows that the initial
volumes of cold and hot gas play a non-negligible role in the functioning of the engine. Observe
that there is a difference up to $20\%$ between the maximum and minimum relative work for
$\theta_0=1.46~\pi$ and $\theta_0=0.46~\pi$. It is clear that the maximum work is obtained when
initially all the gas is in the cold zone, and the minimum work is obtained when initially almost
all the gas is in the hot zone.

We have also integrated numerically the heat equation (Eq.~(22) of Ref.~\onlinecite{Romanelli1}) as
in previous works to obtain the absorbed heat $Q_{in}$ and rejected heat $Q_{out}$, in a
cycle.\cite{Romanelli1, Romanelli2}
\begin{figure}[ht]
\begin{center}
\includegraphics[scale=0.35]{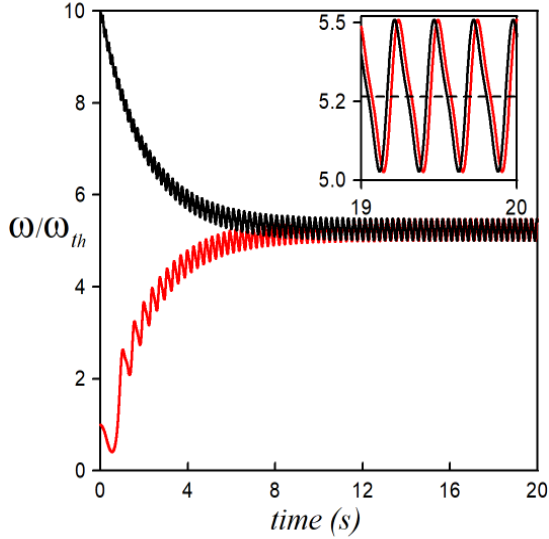}
\end{center}
\caption{Dimensionless angular velocity as a function of time. Initial conditions are $\omega_0=10~
\omega_{th}$ ($\omega_{th}=3\pi{b}/{I}$) for the black curve and $\omega_0=\omega_{th}$  for the
red curve, and in both cases $\theta_0=1.46~\pi$. The inset shows a magnification of both curves
when the steady states are already reached. It is seen that
$\langle\omega\rangle=5.24~\omega_{th}$, (dashed line), and that both curves have the same
asymptotic period, $\mathcal{T'}\simeq~0.25$ seconds.
The parameters are:
$\phi=0.5~\pi$, $\alpha=1.25$, $\beta=1.23$, $a/A=0.016$, $z=0.067$, $b/I=0.5~s^{-1}$ and
$I=1.2~{10}^{-4}kg~m^2$.} \label{f4}
\end{figure}
\section{Dynamics of LTD Stirling engine}
In order to simplify the mechanical model we neglect the masses of piston, displacer and rods. The
conservation of energy imposes that all the power generated by the gas must be transferred to the
the shaft, this means that
\begin{equation}
(P-P_0)\frac{dV}{dt}=\tau \frac{d\theta}{dt}, \label{pote}
\end{equation}
where $\tau$ is the external torque over the shaft and, ${d\theta}/{dt}$ is the angular velocity of
the flywheel. Then $\tau$ is given by
\begin{equation}
\tau=(P-P_0)\frac{dV}{d\theta}. \label{torque}
\end{equation}
The inevitable mechanical friction of the system, as well as a possible additional load
incorporated to extract energy from it, generate an additional friction torque on the shaft, modeled
as $-b~{d\theta}/{dt}$, where  the sign indicates loss of power and $b$ is the damping coefficient.
\begin{figure}[ht]
\begin{center}
\includegraphics[scale=0.35]{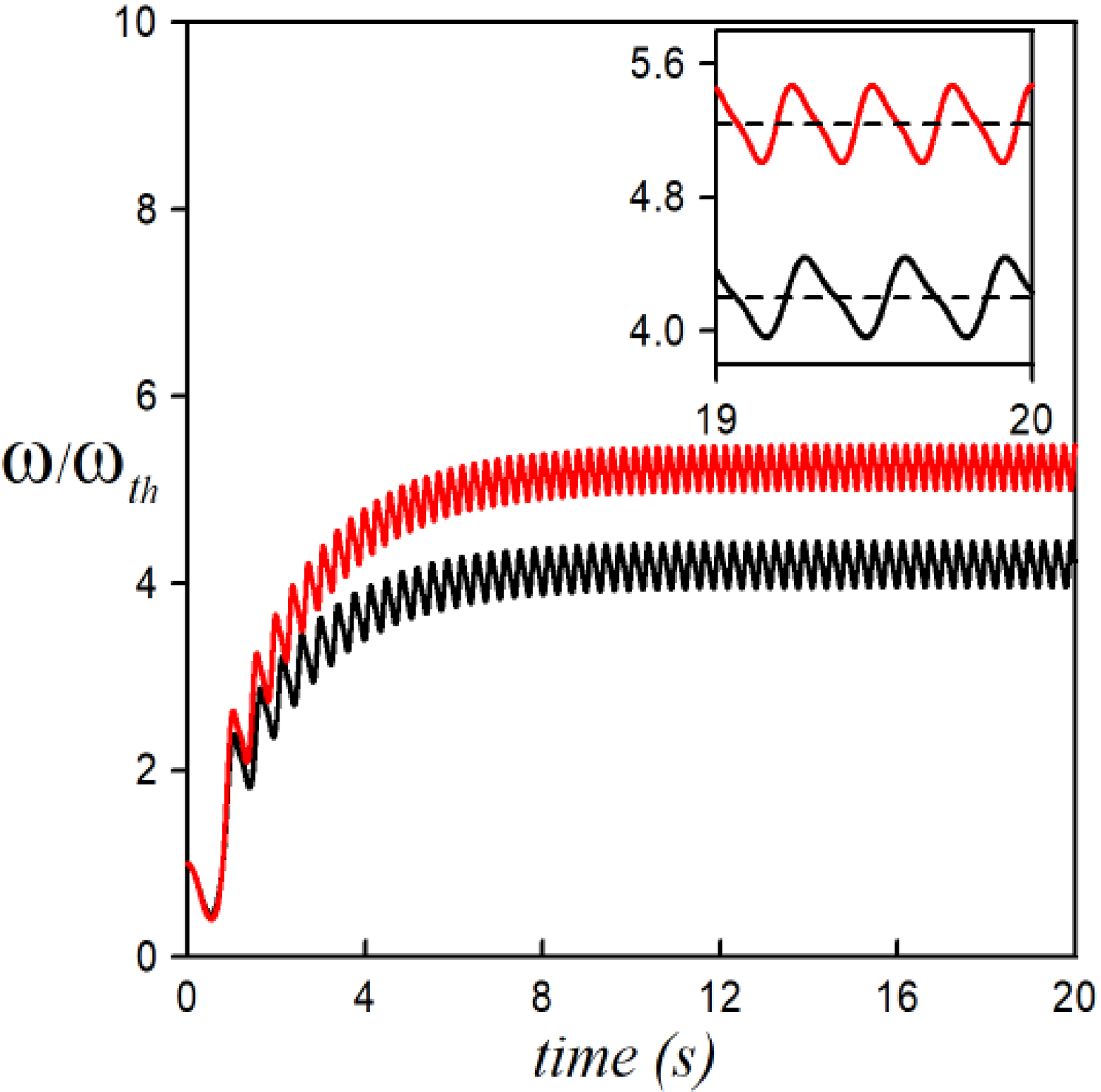}
\end{center}
\caption{Dimensionless angular velocity as a function of time. Initial conditions are
$\theta_0=0.46~\pi$ for the black curve, $\theta_0=1.46~\pi$ for the red curve and for both curves,
$\omega_0=\omega_{th}$. The inset shows a magnification of both curves when the steady states are
already reached. It is seen that for the black curve $\langle\omega\rangle=4.20~\omega_{th}$
(dashed line) and the asymptotic period is $\mathcal{T'}\simeq~0.33$ s. Similarly for the red curve
$\langle\omega\rangle=5.24~\omega_{th}$ and $\mathcal{T'}\simeq~0.25$ s. The values of the
parameters are the same as in Fig.~\ref{f4}.} \label{f5}
\end{figure}
\begin{figure}[ht]
\begin{center}
\includegraphics[scale=0.35]{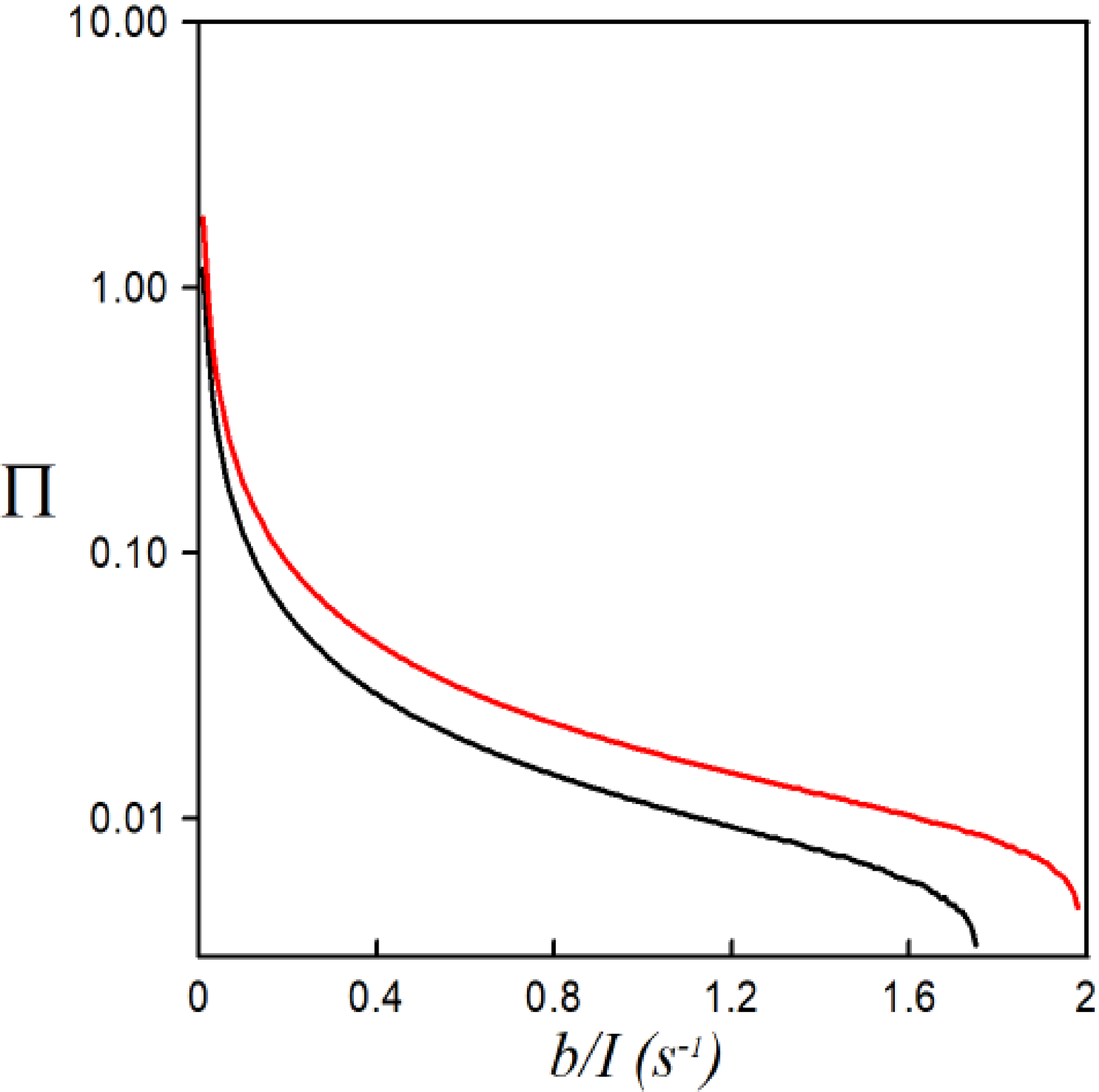}
\end{center}
\caption{Dimensionless gas power (in a logarithmic scale) as a function of a generalized friction
coefficient. For the upper red curve $\theta_0=1.46~\pi$, and for the lower black curve
$\theta_0=0.46~\pi$, and for both curves $\omega_0=40~\omega_{th}$.} \label{f6}
\end{figure}
Therefore the rotational equation of motion indicates that the net external torque on the shaft
determines the rate of change of its angular momentum, \emph{i.e.}
\begin{equation}
I\frac{d^{2}\theta}{dt^{2}}=-b\frac{d\theta}{dt}+(P-P_0)\frac{dV}{d\theta}, \label{newton}
\end{equation}
where $I$ is the flywheel moment of inertia. In order to solve Eq.~(\ref{newton}) we need $P$ and
$V$ as functions of $\theta$ given by Eqs.~(\ref{v}) and (\ref{pe}); as a result we get a nonlinear equation of $\theta$.

Multiplying Eq.~(\ref{newton}) by the angular velocity $\omega\equiv{d\theta}/{dt}$ and integrating
the resulting equation from $t=0$ to the arbitrary time $t$ we obtain the energy equation:
\begin{equation}
\frac{1}{2}I\omega^2-\frac{1}{2}I{\omega_0}^2=-b\int_{0}^{t}{\omega}^2\,dt^{'}+
\int_{\theta_0}^{\theta}(P-P_0){\frac{dV}{d\theta^{'}}}\,d\theta^{'}, \label{energy}
\end{equation}
where $\omega_0$ is the initial angular velocity.
The first term on the right-hand side is the dissipative work and the second is associated with the
engine power. From Eq.~(\ref{pe}) together with Eqs.~(\ref{v2g}) and (\ref{v1g}) it is seen that
$P$ is a periodic function of $\theta$, then the second term of the right-hand side satisfies
\begin{equation}
\int_{\theta_0}^{\theta}(P-P_0){\frac{dV}{d\theta}}\,d\theta=
n\int_{0}^{2\pi}P{\frac{dV}{d\theta}}\,d\theta+
\int_{\theta_0}^{\theta_r}(P-P_0){\frac{dV}{d\theta}}\,d\theta, \label{ww}
\end{equation}
where $n=(\theta-\theta_0)[mod~2\pi]$ is the number of complete cycles and
$\theta_r=(\theta-\theta_0)-2\pi n$.

Equations~(\ref{energy}) and (\ref{ww}) predict that if $b=0$ the kinetic energy grows with the
number of turns $n$, however in real situations we always have $b\neq0$. As it will be shown later
our model predicts that $\omega$ has a finite asymptotic average value, with a little oscillation
$\Delta\omega$ around it.

Dividing both sides of Eq.~(\ref{energy}) by $t$ and taking the limit for $t\rightarrow\infty$ we
obtain the relation between the dissipated power by the friction  and power delivered by the gas
\begin{equation}
b\langle\omega^2\rangle=\frac{1}{\mathcal{T}}\int_{0}^{2\pi}P{\frac{dV}{d\theta}}\,d\theta,
\label{wm}
\end{equation}
where the flywheel period $\mathcal{T}$ is defined as
\begin{align}
\mathcal{T}\equiv &\lim{}\frac{t}{n} \cr
 &t\rightarrow \infty \,   \label{asym0}
\end{align}
and the mean quadratic angular velocity is defined as
\begin{align}
\langle\omega^2\rangle\equiv&\lim{}{\frac{1}{t}}{\int_{0}^{t}{\omega}^2\,dt} .\cr &t\rightarrow
\infty \,   \label{asym}
\end{align}
Finally, using Eq.~(\ref{wm}) we can approximate the asymptotic average value of $\omega$,
$\langle\omega\rangle\equiv{2\pi}/{\mathcal{T}}$, as
\begin{equation}
\langle\omega\rangle\simeq\frac{1}{2\pi b}\int_{0}^{2\pi}P{\frac{dV}{d\theta}}\,d\theta,
\label{wm2}
\end{equation}
if $|\Delta\omega/\langle\omega\rangle|\ll1$.

As mentioned, it is necessary to start the engine with an initial angular velocity $\omega_0$ to
overcome the initial static friction and the dynamic friction that follows it. This velocity must
exceed a certain threshold value $\omega_{th}$ that depends on the nonlinear dynamics of
Eq.~(\ref{newton}) which in turn depends on the parameters $I$, $b$, $\alpha$, $\beta$, $z$, $a/A$
and $\theta_0$. Considering the order of magnitude of the parameters used in this paper, we
estimate a practical value of $\omega_{th}$ solving Eq.~(\ref{newton}) with $P-P_0=0$. In this
case, the solution of Eq.~(\ref{newton}) is
\begin{equation}
\theta=\omega_0\frac{I}{b}(1-e^{-\frac{b}{I}t}). \label{theta}
\end{equation}
We choose arbitrarily $\theta=3\pi$ in order that the flywheel can asymptotically complete at least
one and half cycles. This leads to
\begin{equation}
\omega_{th}\sim\omega_{0}=3\pi\frac{b}{I}\ll\langle\omega\rangle. \label{threshold}
\end{equation}

Now we study numerically the dynamics of the shaft, and in particular its dependence on both the
engine parameters and the initial conditions. Using a standard Runge-Kutta method for the
approximate solutions of ordinary differential equations, we solved Eq.~(\ref{newton}) with the
constraints Eqs.~(\ref{v}) and (\ref{pe}).

\begin{figure}[ht]
\begin{center}
\includegraphics[scale=0.35]{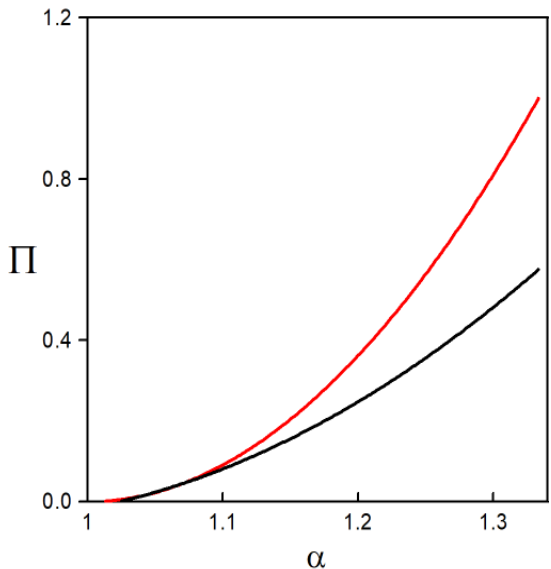}
\end{center}
\caption{Dimensionless gas power as a function of the temperature ratio of heat reservoirs
$\alpha$. For $\alpha=T_h/T_c\sim1.34$  the thermal difference $T_h-T_c\sim100~^0C$. For the upper
red curve $\theta_0=1.46~\pi$, and for the lower black curve $\theta_0=0.46~\pi$, and for both
curves $\omega_0=40~\omega_{th}$.} \label{f7}
\end{figure}

Figures~\ref{f4} and ~\ref{f5} show the flywheel angular velocity as a function of time for two
different initial conditions. Both figures show that the steady state is reached for a sufficiently
large time. These asymptotic states are characterized by a constant average value with a fast small
oscillation around it.  Therefore our toy model predicts an asymptotic time dependent angular
velocity with a well defined period $\mathcal{T'}$, which, in principle, could be different from
$\mathcal{T}$, the flywheel period.

In the case of Fig.~\ref{f4} we use the same value of $\theta_0$ but with two different initial
angular velocities $\omega_0$. This figure shows that in the steady state the angular velocity does
not keep any correlation with its initial value; such a behavior is characteristic of dissipative
systems. In the inset we see that the period $\mathcal{T'}$ is the same for both curves, that is
independent of $\omega_0$.

In Fig.~\ref{f5} we keep the same value for $\omega_0$ but with two different values for
$\theta_0$. Now the asymptotic average angular velocities have different values, this means that
they depend on $\theta_0$, and the same is true for the periods $\mathcal{T'}$.

At this point we emphasize that the numerical calculation shows that the flywheel period
$\mathcal{T}={2\pi}/{\langle\omega\rangle}$ and  the asymptotic period $\mathcal{T'}$ of
$\omega(t)$
 coincide numerically, that is
\begin{equation}
\mathcal{T}=\mathcal{T'}. \label{times}
\end{equation}
This coincidence has no obvious theoretical explanation, all the more so since the dynamical
differential equation is nonlinear. In order to show that the relation given by Eq.~(\ref{times})
could be different, we reason as follows: When the system reaches the steady state, Figs.~\ref{f4}
and~\ref{f5} suggest that $\omega(t)$ can be approximated by
\begin{equation}
\omega(t)=\langle\omega\rangle+\zeta\cos{\frac{2\pi}{\mathcal{T'}}t}, \label{wmdt}
\end{equation}
where $\zeta$ is a small amplitude. However, since the flywheel angular position $\theta(t)$ is
characterized by the period $\mathcal{T}$, the angular velocity (that satisfies $\omega(t)\equiv
d\theta/dt$) must have also the same period. Therefore, $\mathcal{T}$ and $\mathcal{T'}$ must satisfy
the relation
\begin{equation}
\mathcal{T}= m \mathcal{T'}, \label{tytp}
\end{equation}
where $m$ is an integer. Our numerical calculation indicates that $m=1$ but we have no additional
theoretical argument for this value.

Returning to our central development, let us define the power of the engine $\Pi$ by the right-hand
side of Eq.~(\ref{wm}), \emph{i.e.}
\begin{equation}
\Pi\equiv\frac{1}{\mathcal{T}}\int_{0}^{2\pi}P{\frac{dV}{d\theta}}\,d\theta, \label{power}
\end{equation}
of course it is also true that $\Pi=b\langle\omega^2\rangle$.

In Fig.~\ref{f6} we show $\Pi$ as a function of $b/I$ for the initial angular conditions associated
with the maximum and minimum work. The upper limit of friction (\emph{i.e.} the maximum value of
$b/I$) is determined by the vanishing of the angular velocity, hence the vanishing of $\Pi$. The
minimum value of $b/I$ has been chosen to ensure an adequate numerical convergence because
asymptotically $\omega\rightarrow\infty$, the period $\mathcal{T}\rightarrow0$ and
$\Pi\rightarrow\infty$. The figure also shows that the power for $\theta_0=1.46~\pi$ is larger than
for $\theta_0=0.46~\pi$, independently of $b/I$, as it is the case for the work.

Figure~\ref{f7} shows the growth of $\Pi$ with the the temperature ratio $\alpha$ of the heat
reservoirs. It also shows the dependence of $\Pi$ on the initial condition $\theta_0$.

\section{Conclusions}
This paper developed the dynamics and thermodynamics for a Stirling engine that operates at low
temperature difference. The working gas pressure is expressed analytically using an alternative
thermodynamic model developed in a previous paper. This pressure depends on the hot and cold gas
volumes as well as their initial conditions and  the temperature ratio of the heat reservoirs.

The rotational dynamical equation for the shaft includes terms related to the energy delivered by
the engine and to the energy dissipated by friction. This nonlinear equation, which we solved
numerically,  has an explicit dependence on the initial gas volumes through the initial angular
position $\theta_0$.

We showed numerically that both the maximum shaft work and power are obtained when at the start all
the gas is in the cold zone. Conversely the minimum shaft work and power are obtained when
initially almost all the gas is in the hot zone. These results are independent of: (a) the shaft
initial angular velocity, (b) the shaft friction intensity,  (c) the temperature ratio of the heat
reservoirs.

In order to choose initial conditions we estimated a minimal initial flywheel speed to start the
engine. Our model predicts that the flywheel velocity achieves a finite asymptotic average value
with a small periodic oscillation around it. Our system has a dissipative dynamics, however its
steady state showed an unexpected dependence on the initial angular position. This means that the
angular velocity evolves to an attractor independently of its initial value but dependent on the
initial angular position. This asymmetric behavior is due the fact that the model incorporated the
initial angular position in the dynamical equation but not the initial angular velocity.

Summarizing, this paper describes the operation of the LTD Stirling engine in a simple and
understandable way, highlighting some unknown aspects of its behavior.


\section*{ACKNOWLEDGMENTS}
I acknowledge the stimulating discussions with V\'{\i}ctor Micenmacher, and the support from ANII
and PEDECIBA (Uruguay).

\end{document}